\documentclass[preprint,prd,nofootinbib,tightenlines,amsmath,amssymb]{revtex4}
\usepackage{graphicx}
\usepackage{bm}
\usepackage{amsmath}
\usepackage{hhline}
\usepackage{color}
\usepackage{xcolor}

\newcommand{\UBL}{\text{U}(1)_{\text{B}-\text{L}}}
\newcommand{\ZZ}{\mathbb{Z}_2\hspace{-0.05cm}}
\newcommand{\ZZZ}{\mathbb{Z}_3\hspace{-0.05cm}}
\newcommand{\ZN}{\mathbb{Z}_\text{N}\hspace{-0.05cm}}
\newcommand{\EL}{\overline{E}}
\newcommand{\etat}{\tilde{\eta}}
\newcommand{\Phit}{\tilde{\Phi}}
\newcommand{\NR}{N_{\hspace{-0.05cm}R}}
\newcommand{\NRc}{N_{\hspace{-0.05cm}R}^\text{c}}
\newcommand{\NRcbar}{\overline{N_{\hspace{-0.05cm}R}^\text{c}}}
\newcommand{\SigmaR}{\Sigma_{\hspace{-0.02cm}R}}
\newcommand{\SigmaRc}{\Sigma_{\hspace{-0.02cm}R}^\text{c}}
\newcommand{\SigmaRbar}{\overline{\Sigma_{\hspace{-0.02cm}R}}}
\newcommand{\SigmaL}{\Sigma_{\hspace{-0.01cm}L}}
\newcommand{\SigmaLbar}{\overline{\Sigma_{\hspace{-0.01cm}L}}}
\newcommand{\psiR}{\psi_{\hspace{-0.02cm}R}}
\newcommand{\psiRc}{\psi_{\hspace{-0.02cm}R}^\text{c}}
\newcommand{\psiRbar}{\overline{\psi_{\hspace{-0.02cm}R}}}
\newcommand{\psiRcbar}{\overline{\psi_{\hspace{-0.02cm}R}^\text{c}}}
\newcommand{\psiL}{\psi_{\hspace{-0.01cm}L}}

\newcommand{\psiLcbar}{\overline{\psi_{\hspace{-0.01cm}L}^\text{c}}}
\newcommand{\QN}{Q_{\hspace{-0.03cm}N}}
\newcommand{\QSigmaL}{Q_{\Sigma_{\hspace{-0.01cm}L}}}
\newcommand{\QSigmaR}{Q_{\Sigma_{\hspace{-0.01cm}R}}}
\newcommand{\QpsiR}{Q_{\psi_{\hspace{-0.01cm}R}}}
\newcommand{\QpsiL}{Q_{\psi_{\hspace{-0.01cm}L}}}

\begin{document}
\baselineskip=15pt \parskip=5pt

\vspace*{3em}

\preprint{LPT-Orsay-16-36}
\preprint{KUNS-2621}

\title{Systematic $\UBL$ Extensions of Loop-Induced \\ Neutrino Mass Models with Dark Matter} 

\author{
Shu-Yu Ho,$^{1,}$\footnote[1]{sho3@caltech.edu}
Takashi Toma,$^{2,}$\footnote[2]{takashi.toma@th.u-psud.fr}
and 
Koji Tsumura$^{3,}$\footnote[3]{ko2@gauge.scphys.kyoto-u.ac.jp}}
\affiliation{
${}^{1}$Department of Physics, California Institute of Technology, Pasadena, CA 91125, USA \vspace{3pt} \\
${}^{2}$Laboratoire de Physique Th\'eorique, CNRS,\\ 
Univ. Paris-Sud, Universit\'e Paris-Saclay, 91405 Orsay, France \vspace{3pt} \\
${}^{3}$Department of Physics, Kyoto University, Kyoto 606-8502, Japan\vspace{3ex}}


\begin{abstract}
We study the gauged $\hspace{-0.02cm}\UBL\hspace{-0.03cm}$ extensions of the models for neutrino masses and dark matter. In this class of models, tiny neutrino masses are radiatively induced through the loop diagrams, while the origin of the dark matter stability is guaranteed by the remnant of the gauge symmetry. Depending on how the lepton number conservation is violated, these models are systematically classified. We present complete lists for the one-loop $^{}\ZZ^{}$ and the two-loop $^{}\ZZZ^{}$ 
radiative seesaw models as examples of the classification. The anomaly cancellation conditions in these models are also discussed.
\end{abstract}

\maketitle

\section{Introduction}
The standard model (SM) of particle physics has been established after the discovery of the $125\,\text{GeV}$ Higgs boson~\cite{Aad:2012tfa,Chatrchyan:2012xdj}. Nonetheless, there are still remaining puzzles which are not addressed in the SM. In particular, the smallness of the neutrino masses measured by the neutrino-oscillation experiments~\cite{Wendell:2010md,Abdurashitov:2009tn, Abe:2012tg, Gando:2010aa} and the existence of the cosmic dark matter (DM) inferred from the astronomical observations~\cite{Begeman:1991iy, Massey:2007wb, Randall:2007ph, Harvey:2015hha, Adam:2015rua} are phenomenologically important issues. The simplest solution to explain the tiny neutrino masses is the canonical seesaw mechanism with super-heavy right-handed neutrinos~\cite{Minkowski:1977sc, Yanagida:1979as, GellMann:1980vs}, while its verification by experiments may be difficult. On the other hand, various DM candidates - such as the axion, weakly interacting massive particles, asymmetric DM, strongly interacting massive particles and wimpzilla - have been suggested. However, the scale of the DM mass is unknown and spreads over a very wide range from $10^{-15}\,\text{GeV}$ to $10^{15}\,\text{GeV}$~\cite{Baer:2014eja}.

Models with a radiative neutrino mechanism are among the most economical scenarios which can resolve the above two issues at the same time. In this class of models, the neutrino mass is induced by quantum effects, while the DM candidate is incorporated as a necessary component. Since the DM particle is running in the quantum loop diagram in order to generate the neutrino masses, the phenomenology of neutrinos and that of DM are strongly correlated.
Representative models possessing these ingredients include Ma's scotogenic models at the one-loop~\cite{Ma:2006km}, and the two-loop level~\cite{Ma:2007gq}, the Krauss-Nasri-Trodden (KNT) model at the three-loop level~\cite{Krauss:2002px},  
and other three-loop models proposed by Aoki, Kanemura, and Seto ~\cite{Aoki:2008av}, and by Gustafsson, No, and Rivera~\cite{Gustafsson:2012vj}.
In each of these models, an {\em ad hoc} discrete symmetry $\ZZ$ or $\ZZZ$ 
is imposed not only to forbid the tree-level neutrino Yukawa interactions, but also to guarantee the DM stability. The origin of the discrete symmetry is left unknown. A possible origin of this symmetry is an accidental symmetry. If one extends the SM with higher-dimensional \,SU(2)$_L$ representations such as quintets or septets, an accidental \,$\ZZ$\, symmetry appears in a new particle sector and stabilizes the DM candidate~\cite{Cirelli:2005uq, Farina:2013mla,
DelNobile:2015bqo}.\footnote{A large isospin {\em scalar} multiplet leads a lower cutoff scale, 
which might disturb the DM stability
~\cite{Hamada:2015bra}.} 
Extensions of the models with radiative neutrino mass generation along this line and their phenomenology have been explored in Refs.~\cite{Ahriche:2015wha, Ahriche:2016rgf, Sierra:2016qfa}. 
Another attractive idea for the dark matter stability is that the discrete symmetry originates from a continuous symmetry which is spontaneously broken at some high energy scale by a nonzero vacuum expectation value (VEV) of scalar fields. If the continuous symmetry is the gauge symmetry, this mechanism is known as the Krauss-Wilczek mechanism~\cite{Krauss:1988zc}. This mechanism has been applied to the radiative seesaw models~\cite{Chang:2011kv}. 
The spontaneous breaking of a global symmetry can also leave a $\,\ZZ\,$ symmetry, which is well known as the domain wall production in axion models. This residual symmetry has also been used to construct a radiative seesaw model~\cite{Dasgupta:2013cwa}.

In this paper, we study the gauged $\hspace{-0.03cm}\UBL\hspace{-0.08cm}$ extension of the radiative seesaw models for neutrino masses. The $\ZN$ discrete symmetry can be realized as a remnant of the $\UBL$ gauge symmetry. 
Although this kind of models have been studied in the literature~\cite{Kanemura:2014rpa,Chiang:2013kqa, Ma:2015mjd}, 
our aim is to classify these models systematically. By focusing on how the lepton number conservation is broken in the Feynman diagrams for the neutrino mass generation, we present a list of all possible models. We do not give a detailed numerical analysis of each model at this stage. In general, these $\UBL$ extended models encounter the gauge anomaly problem. A systematic method of anomaly cancellations involving adding extra fermions is discussed.
%

The structure of the paper is as follows. In the next section, we present a systematic procedure for classifying the models for the radiative neutrino mass generation and for the DM stability based on the gauged $\UBL$ symmetry. 
In Sec.~\ref{sec:3}, we deal with the gauge anomaly cancellations. The required number of the extra fermions and their B-L charges are given in the Appendix as examples. We conclude and summarize our study in Sec.~\ref{sec:5}.

\section{$\UBL$ Extensions}
\label{sec:2}
We demonstrate the $\UBL\hspace{-0.02cm}$ extension of the radiative seesaw models at the one-loop, two-loop, and three-loop levels~\cite{Ma:2006km,Ma:2007gq, Krauss:2002px}. 
A discrete $\ZN$ symmetry is derived as a residual symmetry of the $\UBL$ gauge symmetry. 
In the following investigation, the (minimal) models include gauge anomalies in general. For the moment, we ignore the issue of the anomaly cancellations. These anomalies can be canceled by introducing vector-like fermions (under the SM gauge group)~\cite{Appelquist:2002mw, Batra:2005rh, Kanemura:2011mw}. 
The details of the systematic cancellation of gauge anomalies will be addressed in the next section.

\subsection{One-loop $\ZZ$ Model}
First, we consider the $\UBL\hspace{-0.03cm}$ extension of the one-loop model with the $\ZZ$ symmetry~\cite{Ma:2006km}.\footnote{The phenomenology of this model has been studied, for example, in Refs.~\cite{Kubo:2006yx, Suematsu:2009ww, Schmidt:2012yg, Ho:2013hia, Ho:2013spa, Faisel:2014gda} for example.}  
In the original non-gauged model, three right-handed singlet fermions $\NR$ and 
one inert doublet scalar $\eta=(\eta^+,\eta^0)^\text{T}\hspace{-0.08cm}$ are added to the SM. 
In addition, the $\ZZ$ parity is assigned as odd for the new particles and even for the SM particles. The lightest electrically neutral $\ZZ$ odd particle, which is either the lightest right-handed fermion or the neutral component of the inert scalar, can be a DM candidate. At least two generations of the right-handed fermions are needed to fit the observed neutrino masses and mixings. The required interactions for generating neutrino masses are written down as
\begin{eqnarray}\label{eq:ma}
\mathcal{L} \,\, \supset \,\,
\EL \NR \hspace{0.02cm} \etat ~, \quad
\NRcbar \hspace{0.02cm} \NR  ~, \quad
\big(\Phi^\dag\eta\big)^{\hspace{-0.05cm}2}  ~,~
+\,\text{h.c.} ~\,\,, \quad
\end{eqnarray}
where $E$ $(\Phi)$ is the SM lepton (Higgs) doublet, $\etat\,=\,i\tau_2\eta^\ast$ with 
$\tau_2$ being the usual second Pauli matrix, and 
$\NRc\,\equiv(\NR)^\text{c}$ denotes the charge conjugate of $\NR$\,.

\begin{table}[t]
\begin{center}
\begin{tabular}
[c]{|c||c|c||c|c|c|c|}\hline
~$\vphantom{|_|^|}$~ & ~$E$~ &  ~$\Phi$~ & ~$\NRc$~ & ~$\eta$~ & ~$\chi$~ & ~$\sigma$~ 
\\\hline\hline
~SU(2)$_L\vphantom{|_|^|}$~   
& ~$\bm{2}$~ & ~$\bm{2}$~ & ~$\bm{1}$~ & ~$\bm{2}$~ & ~$\bm{1}$~ & ~$\bm{1}$~ 
\\\hline
~U(1)$_{Y}\vphantom{|_|^|}$~ 
& ~$-$1/2~ & ~1/2~ & ~0~ & ~1/2~ & ~0~ & ~0~                                                              
\\\hline
~U(1)$_{\text{B}-\text{L}}\vphantom{|_|^|}$~ 
& ~$-$1~ & ~0~ & ~$\QN$~ & ~$Q_\eta$~ & ~$Q_\chi$~ & ~$Q_\sigma$~
\\\hline
~Spin~$J\vphantom{|_|^|}$~ & ~1/2~ & ~0~ & ~1/2~ & ~0~ & ~0~ & ~0~
\\\hline
\end{tabular}
\caption{
Charge assignments of the fermions and scalars in the $\UBL\hspace{-0.03cm}$ model 
with one-loop neutrino mass generation, where $\QN$, $Q_\eta$, $Q_\chi$ and 
$Q_\sigma$ $(Q_{\chi}\,,Q_{\sigma} \neq 0)$ are determined properly as discussed in the context.}
\label{tab:1}
\end{center}
\vspace{-0.3cm}
\end{table}

In order to achieve the $\UBL$ extension, 
two new complex singlet scalars $\chi$ and $\sigma$ are added to the original one-loop model. 
The particle contents and the charge assignments are shown in Table~\ref{tab:1}. 
%
The B$-$L charges of quarks and of right-handed charged leptons are not displayed in Table~\ref{tab:1};
these are fixed appropriately as usual. 
Hereafter, we assume that among the new scalar fields only the $\sigma$ field develops a VEV, 
which triggers the $\UBL$ symmetry breaking. 
The third term in Eq.\,\eqref{eq:ma} is not allowed in the $\UBL\hspace{-0.03cm}$ extended model 
because $\Phi$ must be neutral while $\eta$ should be charged under the $\UBL$ symmetry. 
In order to effectively induce this term, we need a mixing between $\eta^0$ and $\chi$ when the $\UBL$ symmetry is broken. 
In this extended setup, the necessary interactions for the radiative seesaw mechanism are
\begin{eqnarray}\label{eq:ma_ex}
\mathcal{L} \,\, \supset 
\begin{array}{c|l|ll|l}
& \NRcbar \hspace{0.02cm} \NR
& (\Phi^\dag \eta) \chi & (\Phi^\dag \eta) \chi^\ast
& \chi^2 \sigma \\
\EL \NR \hspace{0.02cm} \etat
& \NRcbar \hspace{0.02cm} \NR \hspace{0.03cm} \sigma  
& (\Phi^\dag \eta) \chi \sigma 
& (\Phi^\dag \eta) \chi^\ast \sigma 
&  \chi^2 \sigma^\ast \\
& \NRcbar \hspace{0.02cm} \NR \hspace{0.03cm} \sigma^\ast
& (\Phi^\dag \eta) \chi \sigma^\ast
& (\Phi^\dag \eta) \chi^\ast \sigma^\ast \\
\end{array}
\hspace{-0.05cm}+\,\text{h.c.} ~.
\end{eqnarray}
At least one element in each column has to be selected in an extended model. Since these terms must be invariant under the $\UBL\hspace{-0.07cm}$ transformation, 
the unknown charges $(\QN\,,Q_\eta\,,Q_\chi\,,Q_\sigma)$ are determined by solving the simultaneous equations. Once all the sets of the B$-$L charges are found, the entire Lagrangian can be easily constructed. Notice that the requirement of the scalar interactions in the last column of Eq.\,\eqref{eq:ma_ex} 
forbids the $\chi$ linear terms, $\sigma^n\chi$ and $(\sigma^\ast)^n\chi$ $(n=1,2,3)$, 
which causes the tadpole diagram (or induced VEV) to conflict with the remnant $\ZZ$ symmetry. 

\begin{figure}[b]
\vspace{-0.1cm}
\begin{center}
\includegraphics[scale=0.26]{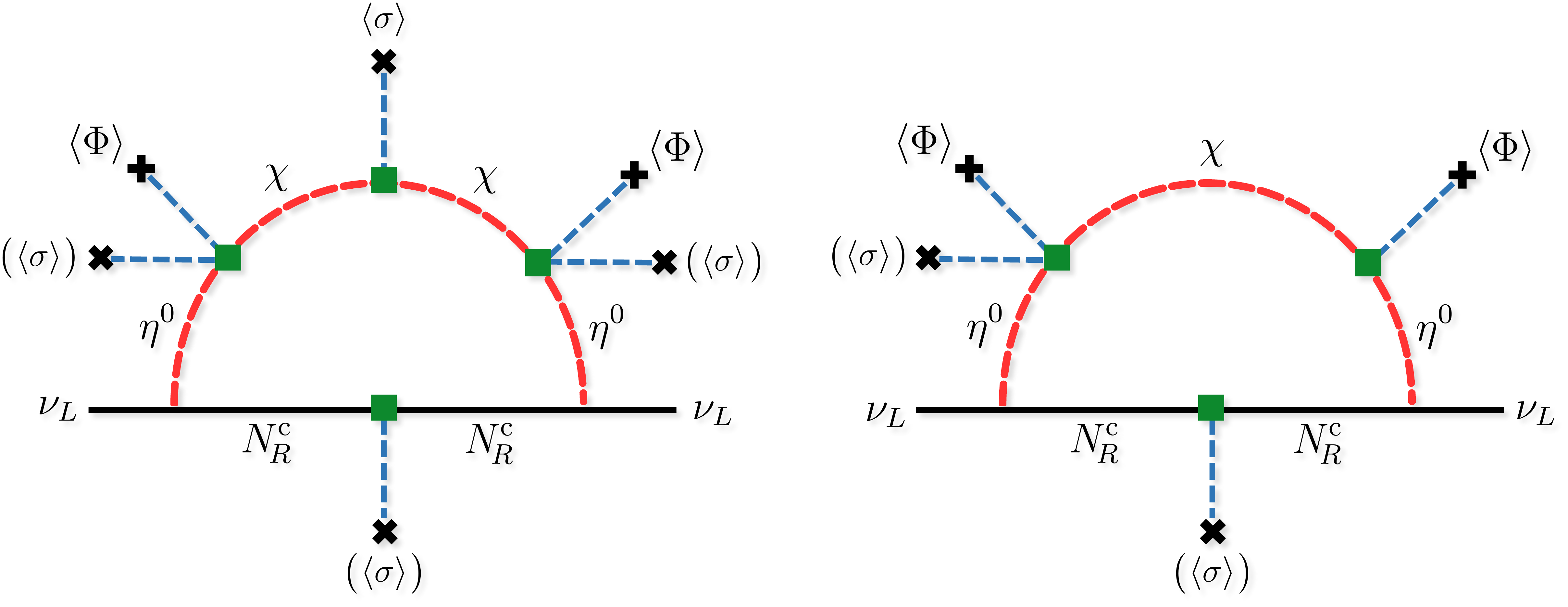}
\vspace{-0.3cm}
\caption{Feynman diagrams for the neutrino mass generation in the $\UBL\hspace{-0.03cm}$ extension of the one-loop model, where the squares indicate the possible B$-$L breaking vertex.}
\label{fig:one_loop}
\end{center}
\vspace{-0.5cm}
\end{figure}

The topological diagrams for the one-loop neutrino mass generation are shown in Fig.~\ref{fig:one_loop}, 
where the square symbol indicates the possible B$-$L breaking vertex. 
Depending on the B$-$L charge assignment as discussed above, 
the combinations of the lepton-number-violating vertices change 
In an opposite way, we can firstly identify the possible B$-$L breaking vertices in the diagram. 
In each diagram, the sum of the B$-$L violation is limited to two since the Majorana mass terms break  
the lepton number by two units. 
Therefore, the combinations of these vertices fix all the $\UBL\hspace{-0.03cm}$ charges of new particles. 
%
%
The possible B$-$L charge assignments are summarized in Table~\ref{tab:2}. 
The assignments obtained by the field redefinitions $\chi^\ast\to\chi$ \,and/or\, $\sigma^\ast\to\sigma$ 
are regarded as the same model.  
The assignments A1 and A4 involve two kinds of terms, which induce the mixing between $\eta^0$ and $\chi$\,, 
namely, $(\Phi^\dag\eta)\chi^\ast$ and $(\Phi^\dag\eta)\chi^{}\sigma$. 
The other assignments include only one mixing term. 
We again note that the nature of the lepton number violation is different for each assignment. For instance, in the case of A1, A2, and A3, since all the new particles carry nonzero B$-$L charge, both of the fermion and scalar lines in Fig.~\ref{fig:one_loop} violate the lepton number. 
On the other hand, for the assignments A4 and A5, since the new fermion $\NRc$ has no B$-$L charge, the lepton number violation occurs only in the scalar sector.

\begin{table}[t]
\begin{center}
\begin{tabular}[c]{|c||c|c|c|c
||c|}\hline
$\vphantom{|_|^|}$  & $\QN$  & $Q_\eta$ & $Q_\chi$ & $Q_\sigma$ 
& ~necessary interactions~ 
\\\hline\hline
~A1$\vphantom{|_|^|}$~
& ~1/2~    & ~1/2~    & ~1/2~   & ~$-$1~       
& 
~$
\EL \NR \hspace{0.02cm} \etat ~,~
\NRcbar \hspace{0.02cm} \NR \hspace{0.03cm} \sigma^\ast ~,~
(\Phi^\dag\eta) \chi^\ast ~,~
(\Phi^\dag\eta) \chi^{} \sigma ~,~
\chi^2 \sigma
$~     
\\\hline
~A2$\vphantom{|_|^|}$~
& ~1/4~    & ~3/4~    & ~1/4~        & ~$-$1/2~    
&
~$
\EL \NR \hspace{0.02cm} \etat ~,~
\NRcbar \hspace{0.02cm} \NR \hspace{0.03cm} \sigma^\ast ~,~
(\Phi^\dag\eta) \chi^\ast \sigma ~,~
\chi^2 \sigma
$~     
\\\hline
~A3$\vphantom{|_|^|}$~
& ~$-$1/2~    & ~3/2~    & ~$-$1/2~   & ~$1$~        
&
~$
\EL \NR \hspace{0.02cm} \etat ~,~
\NRcbar \hspace{0.02cm} \NR \hspace{0.03cm} \sigma^\ast ~,~
(\Phi^\dag\eta) \chi \sigma^\ast ~,~
\chi^2 \sigma
$~     
\\\hline
~A4$\vphantom{|_|^|}$~
& ~0~      & ~1~       & ~1  ~    & ~$-$2~       
&
~$
\EL \NR \hspace{0.02cm} \etat ~,~
\NRcbar \hspace{0.02cm} \NR ~,~
(\Phi^\dag\eta) \chi^\ast ~,~
(\Phi^\dag\eta) \chi^{}\sigma ~,~
\chi^2 \sigma
$~     
\\\hline
~A5$\vphantom{|_|^|}$~
& ~0~     & ~1~      & ~1/3~      & ~$-$2/3~    
&
~$
\EL \NR \hspace{0.02cm} \etat ~,~
\NRcbar \hspace{0.02cm} \NR ~,~
(\Phi^\dag\eta) \chi^\ast \sigma ~,~
\chi^2 \sigma
$~     
\\\hline
\end{tabular}
\caption{
Possible B$-$L charge assignments for new particles in the $\UBL$ extended one-loop model. 
In each of these assignments, the last column contains the necessary interactions for generating neutrino masses at the one-loop level.}
\label{tab:2}
\end{center}
\vspace{-0.3cm}
\end{table}


For the assignments A1-A3, the remnant discrete symmetry 
$\ZZ\hspace{0.05cm}\equiv(-1)^{\text{N}Q_{\text{B}-\text{L}}}\hspace{-0.06cm}$ is identical to the one in the original non-gauged model~\cite{Ma:2006km} when the $\UBL\hspace{-0.03cm}$ symmetry is broken, 
where $\text{N}Q_{\text{B}-\text{L}}\,(\,\equiv\hspace{-0.02cm}
2\hspace{0.02cm}Q_{\text{B}-\text{L}}/Q_\sigma^{})$.\footnote{The
prefactor $2$ in the definition $\text{N}Q_{\text{B}-\text{L}}$ is
fixed depending on the remnant $\mathbb{Z}_\text{N}$ symmetry. As we
will see later, $\text{N}Q_{\text{B}-\text{L}}$ is defined by
$3\hspace{0.02cm}Q_{\text{B}-\text{L}}/Q_\sigma^{}$ or 
$4\hspace{0.02cm}Q_{\text{B}-\text{L}}/Q_\sigma^{}$ for the model with 
$\mathbb{Z}_3$ or $^{}\mathbb{Z}_4\hspace{-0.03cm}$ symmetry, respectively.}
%
While for the assignments A4 and A5, $\ZZ\hspace{0.05cm}\equiv({-}1)^{\text{N}Q_{\text{B}-\text{L}}+2J}$ is the same as the original model, where $J$ is the spin of the particle. This is because $(-1)^{2J}$ is an accidental \,$\ZZ$ symmetry under the Lorentz transformation, and a product of \,$\ZZ$\, symmetries leads to another \,$\ZZ$\, symmetry.
With this definition of the remnant $\mathbb{Z}_2$ symmetry, the $\mathbb{Z}_2$
charges of the new particles are explicitly fixed to be odd for $\NRc$, $\eta$,
$\chi$ and even for $\sigma$.

\subsection{Two-loop $\ZZZ$ Model I}
The two types of two-loop radiative seesaw models with
the $\ZZZ$ symmetry have been proposed in Ref.~\cite{Ma:2007gq}. 
The phenomenology of these models has also been discussed in the literature~\cite{Aoki:2014cja, Ding:2016wbd}. 
In the first two-loop model, a \,SU(2)$_L\hspace{-0.05cm}$ doublet vector-like fermions 
$\Sigma=(\Sigma^+, \Sigma^0)^\text{T}\hspace{-0.08cm}$, a singlet vector-like fermion $\psi^{}$, 
and three complex singlet scalars $\chi$ are added to the SM.\footnote{Conversely, one can add three \,SU(2)$_L\hspace{-0.03cm}$ doublet vector-like fermion, and a singlet vector-like fermion, and one singlet complex scalar to explain the observed neutrino data.} 
Using the cubic root $^{}\omega$ of unity,
the $\,\ZZZ\,$ charge is assigned as \,$\omega\,=\,e^{2\pi i/3}$ \,or \,$\omega^*$ for the new particles, 
and unity for the SM particles. 
The lightest scalar $\chi$ can serve as a DM candidate. 
A fermionic DM candidate given by the lightest mass eigenstate composed of \,$\Sigma^0$ and $\psi$ 
may not be suitable for a DM candidate since the elastic cross section with nuclei via $Z$ boson exchange is strongly constrained by direct detection searches~\cite{Akerib:2015rjg}. 
The required interactions for generating neutrino masses in the first two-loop model are
\begin{eqnarray}
\mathcal{L} \,\, \supset \,\,
\EL \hspace{0.03cm} \SigmaR \hspace{0.03cm} \chi^\ast ~,\quad
\SigmaRbar \hspace{0.03cm} \SigmaL ~,\quad
\psiRbar \hspace{0.03cm} \psiL ~,\quad
\SigmaLbar  \psiR \hspace{0.02cm} \Phit ~,\quad
\psiLcbar \psiL \chi ~,\quad
\chi^3 ~,~
+\,\text{h.c.} ~\,\,. \quad
\end{eqnarray}

\begin{figure}[htbp]
\vspace{0.4cm}
\begin{center}
\includegraphics[scale=0.25]{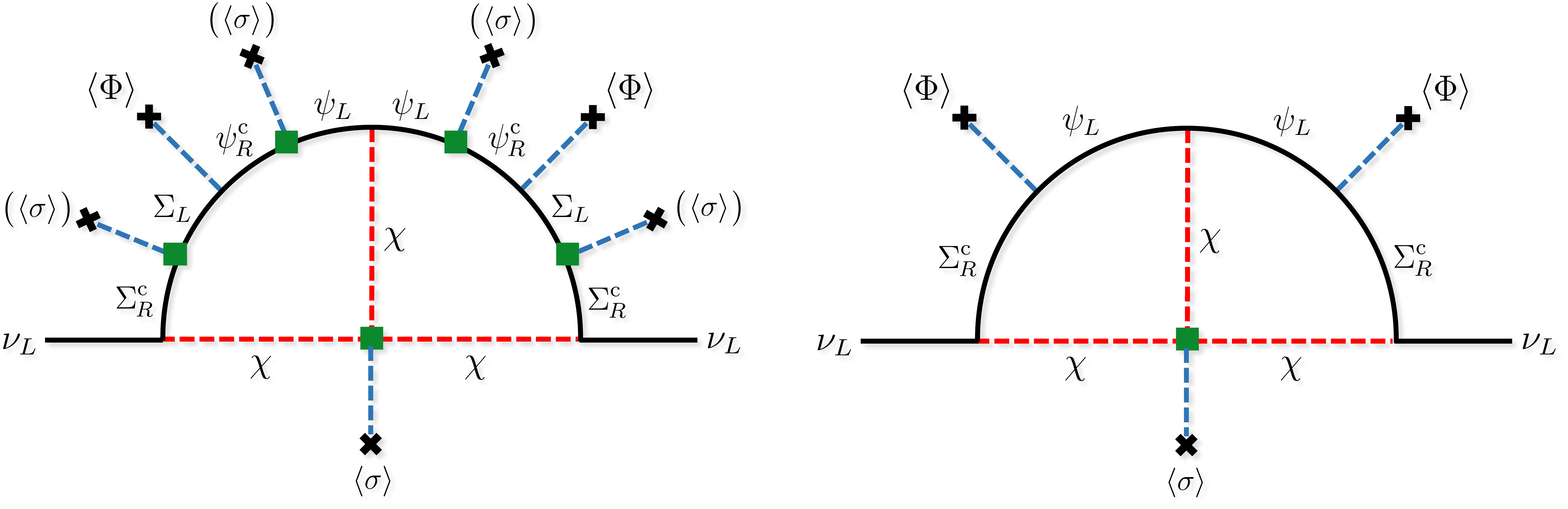}
\vspace{-0.2cm}
\caption{Feynman diagrams for the neutrino mass generation in the $\UBL\hspace{-0.03cm}$ extension of the two-loop model I.}
\label{fig:two_loop_I}
\end{center}
\vspace{-0.3cm}
\end{figure}

To accomplish the $\UBL\hspace{-0.03cm}$ extension, 
we add one singlet complex scalar \,$\sigma$\, to the original model. 
The particle contents and the charge assignment are summarized in Table~\ref{tab:3}.
The necessary interactions to produce neutrino masses in the extended model are given by
%
\begin{eqnarray}
\mathcal{L} \,\, \supset 
\begin{array}{c|l|l|l|ll|l}
& \SigmaRbar \hspace{0.03cm} \SigmaL 
& \psiRbar \hspace{0.03cm} \psiL
& \SigmaLbar \psiR  \hspace{0.02cm} \Phit 
& \psiLcbar \psi_L \chi  
& \psiRcbar \hspace{0.03cm} \psiR \hspace{0.03cm} \chi  
& \chi^3 \sigma \\
\EL \hspace{0.03cm} \SigmaR \hspace{0.03cm} \chi^\ast
& \SigmaRbar \hspace{0.03cm} \Sigma_L \sigma
& \psiRbar \hspace{0.02cm} \psi_L \sigma 
& \SigmaRbar \hspace{0.03cm}  \psiL \Phit
& \psiLcbar \psi_L \chi^\ast
& \psiRcbar \hspace{0.03cm} \psiR \hspace{0.03cm} \chi^\ast  
& \chi^3 \sigma^\ast \\
& \SigmaRbar \hspace{0.03cm} \Sigma_L \sigma^\ast 
& \psiRbar \hspace{0.03cm} \psi_L \sigma^\ast
&&&\\
\end{array}
+\,\text{h.c.} ~.
\end{eqnarray}
At least one element in each column should be chosen to generate neutrino masses at the two-loop level. The topological diagrams for the neutrino mass generation are shown in Fig.~\ref{fig:two_loop_I}. 
By applying the same methodology illustrated in the previous subsection, all the possible charge assignments can be found. These are summarized in Table~\ref{tab:4}. 
There are nine different ways (B1-B9) of assigning charges. In the assignments B5-B7, the Yukawa interaction $ \hspace{0.03cm} \SigmaRbar \hspace{0.03cm}  \psiL \Phit$ 
is also allowed. 
This Yukawa interaction can also be the origin of neutrino masses through the additional diagram in the right panel of Fig.~\ref{fig:two_loop_I}. 
After the $\UBL\hspace{-0.05cm}$ symmetry breaking, the \,$\ZZZ$\, symmetry is kept unbroken. 
The $\ZZZ$ charges for each particle are determined by
$(-1)^{\text{N}Q_{\text{B}-\text{L}}+2J}\hspace{-0.06cm}$.
The resultant charge assignment is consistent with those in the original two-loop $^{}\ZZZ^{}$ model. 
%
\begin{table}[t]
\begin{center}
\begin{tabular}
[c]{|c||c|c||c|c|c|c|c|c|}\hline
~$\vphantom{|_|^|}$~ & ~$E$~ &  ~$\Phi$~ &~$\SigmaL$~ & ~$\SigmaRc$~ & ~$\psiL$~ & ~$\psiRc$~ & ~$\chi$~ & ~$\sigma$~ 
\\\hline\hline
~SU(2)$_L\vphantom{|_|^|}$~   
& ~$\bm{2}$~ & ~$\bm{2}$~ & ~$\bm{2}$~ & $\bm{2}$ & ~$\bm{1}$~ & ~$\bm{1}$~& ~$\bm{1}$~& ~$\bm{1}$~ 
\\\hline
~U(1)$_{Y}\vphantom{|_|^|}$~ 
& ~$-$1/2~ & ~1/2~ & ~$-$1/2~ & ~1/2~ & ~0~ & ~0~ & ~0~ & ~0~ 
\\\hline
~U(1)$_{\text{B}-\text{L}}\vphantom{|_|^|}$~ 
& ~$-$1~ & ~0~ & ~$\QSigmaL$~ & ~$\QSigmaR$~ & ~$\QpsiL$~ & ~$\QpsiR$~ & ~$Q_\chi$~ & ~$Q_\sigma$~
\\\hline
~Spin~$J\vphantom{|_|^|}$~ & ~1/2~ & ~0~ & ~1/2~ & ~1/2~ & ~1/2~& ~1/2~ & ~0~ & ~0~
\\\hline
\end{tabular}
\caption{
Charge assignments of the fermions and scalars in the $\UBL\hspace{-0.03cm}$ extension of the two-loop model I, 
where the unknown B$-$L charges can be fixed by adapting the same strategy as in the previous subsection.}
\label{tab:3}
\vspace{-0.4cm}
\end{center}

\end{table}

\begin{table}[t]
\begin{center}
\begin{tabular}
[c]{|c||c|c|c|c|c|c||c|}\hline
$\vphantom{|_|^|}$   & ~$\QSigmaL$~ & ~$\QSigmaR$~ & ~$\QpsiL$~ & ~$\QpsiR$~ & ~$Q_\chi$~ & ~$Q_\sigma$~           
& ~necessary interactions~
\\\hline
\,B1$\vphantom{|_|^|}$\, & \,$-$7/15\, & \,13/15\, & \,$-$1/15\, & \,7/15\,   & \,2/15\,  & \,$-$2/5\, 
&
~$
\EL \hspace{0.03cm} \SigmaR \hspace{0.03cm} \chi^\ast \,,\,
\SigmaRbar \hspace{0.02cm} \Sigma_L \sigma \,,\, 
\psiRbar \hspace{0.02cm} \psi_L \sigma \,,\,
\SigmaLbar  \psiR \hspace{0.02cm} \Phit \,,\,
\psiLcbar \psi_L \chi \,,\,
\chi^3 \sigma
$~
\\\hline
\,B2$\vphantom{|_|^|}$\, & \,$-$7/9\,   & \,7/9\,   & \,$-$1/9\,   & \,7/9\,     & \,2/9\,    & \,$-$2/3\, 
&
~$
\EL \hspace{0.03cm} \SigmaR \hspace{0.03cm} \chi^\ast \,,\,
\SigmaRbar \hspace{0.02cm} \Sigma_L \,,\,
\psiRbar \hspace{0.02cm} \psi_L \sigma \,,\,
\SigmaLbar  \psiR \hspace{0.02cm} \Phit \,,\,
\psiLcbar \psi_L \chi \,,\,
\chi^3 \sigma
$~
\\\hline
\,B3$\vphantom{|_|^|}$\, & \,$-$1/9\,   & \,7/9\,   & \,$-$1/9\,   & \,1/9\,     & \,2/9\,    & \,$-$2/3\, 
&
~$
\EL \hspace{0.03cm} \SigmaR \hspace{0.03cm} \chi^\ast \,,\,
\SigmaRbar \hspace{0.02cm} \Sigma_L \sigma \,,\,
\psiRbar \hspace{0.02cm} \psi_L \,,\,
\SigmaLbar  \psiR \hspace{0.02cm} \Phit \,,\,
\psiLcbar \psi_L \chi \,,\,
\chi^3 \sigma
$~
\\\hline
\,B4$\vphantom{|_|^|}$\, & \,$-$5/9\,   & \,11/9\,   & \,1/9\,      & \,5/9\,       &  \,$-$2/9\,  & \,$-$2/3\, 
&
~$
\EL \hspace{0.03cm} \SigmaR \hspace{0.03cm} \chi^\ast \,,\,
\SigmaRbar \hspace{0.02cm} \Sigma_L \sigma \,,\,
\psiRbar \hspace{0.02cm} \psi_L \sigma \,,\,
\SigmaLbar  \psiR \hspace{0.02cm} \Phit \,,\,
\psiLcbar \psi_L \chi \,,\,
\chi^3 \sigma^\ast
$~
\\\hline
\,B5$\vphantom{|_|^|}$\, & \,$-$7/3\,   & \,1/3\,    & \,$-$1/3\,   & \,7/3\,     & \,2/3\,        & \,$-$2\,  
&
~$
\EL \hspace{0.03cm} \SigmaR \hspace{0.03cm} \chi^\ast \,,\,
\SigmaRbar \hspace{0.02cm} \Sigma_L \sigma^\ast \,,\,
\psiRbar \hspace{0.02cm} \psi_L \sigma \,,\,
\SigmaLbar  \psiR \hspace{0.02cm} \Phit \,,\,
\psiLcbar \psi_L \chi \,,\,
\chi^3 \sigma
$~
\\\hline
\,B6$\vphantom{|_|^|}$\, & \,5/3\,        & \,1/3\,     & \,$-$1/3\,   & \,$-$5/3\,  & \,2/3\,      & \,$-$2\, 
&
~$
\EL \hspace{0.03cm} \SigmaR \hspace{0.03cm} \chi^\ast \,,\,
\SigmaRbar \hspace{0.02cm} \Sigma_L \sigma \,,\,
\psiRbar \hspace{0.02cm} \psi_L \sigma^\ast \,,\,
\SigmaLbar  \psiR \hspace{0.02cm} \Phit \,,\,
\psiLcbar \psi_L \chi \,,\,
\chi^3 \sigma
$~
\\\hline
\,B7$\vphantom{|_|^|}$\, & \,$-$1/3\,   & \,1/3\,     & \,$-$1/3\,  & \,1/3\,  & \,2/3\,    & \,$-$2\,   
&
~$
\EL \hspace{0.03cm} \SigmaR \hspace{0.03cm} \chi^\ast \,,\,
\SigmaRbar \hspace{0.02cm} \Sigma_L \,,\,
\psiRbar \hspace{0.02cm} \psi_L \,,\,
\SigmaLbar  \psiR \hspace{0.02cm} \Phit \,,\,
\psiLcbar \psi_L \chi \,,\,
\chi^3 \sigma
$~
\\\hline
\,B8$\vphantom{|_|^|}$\, & \,1/3\,   & \,5/3\,    & \,1/3\, & \,$-$1/3\,  & \,$-$2/3\, & \,$-$2\,   
&
~$
\EL \hspace{0.03cm} \SigmaR \hspace{0.03cm} \chi^\ast \,,\,
\SigmaRbar \hspace{0.02cm} \Sigma_L \sigma \,,\,
\psiRbar \hspace{0.02cm} \psi_L \,,\,
\SigmaLbar  \psiR \hspace{0.02cm} \Phit \,,\,
\psiLcbar \psi_L \chi \,,\,
\chi^3 \sigma^\ast
$~
\\\hline
\,B9$\vphantom{|_|^|}$\, & \,$-$5/3\,   & \,5/3\,    & \,1/3\, & \,5/3\,  & \,$-$2/3\,    & \,$-$2\,      
&
~$
\EL \hspace{0.03cm} \SigmaR \hspace{0.03cm} \chi^\ast \,,\,
\SigmaRbar \hspace{0.02cm} \Sigma_L \,,\,
\psiRbar \hspace{0.02cm} \psi_L \sigma \,,\,
\SigmaLbar  \psiR \hspace{0.02cm} \Phit \,,\,
\psiLcbar \psi_L \chi \,,\,
\chi^3 \sigma^\ast
$~
\\\hline
\end{tabular}
\caption{
Possible B$-$L charge assignments for new particles in the two-loop $\UBL\hspace{-0.03cm}$ model I. 
The last column contains the necessary interactions to generate neutrino masses at the two-loop level.
}
\label{tab:4}
\end{center}
\vspace{-0.5cm}
\end{table}

\subsection{Two-loop $\mathbb{Z}_3$ Model II}
For the second two-loop model in Ref.~\cite{Ma:2007gq}, one \,SU(2)$_L$ doublet scalar \,$\eta$\, is added 
instead of a pair of \,SU(2)$_L\hspace{-0.05cm}$ doublet vector-like fermion $\Sigma$ in the first two-loop model. 
The other parts are not different from the first model. The required interaction terms for generating neutrino masses in the second model are
\begin{eqnarray}
\mathcal{L} \,\, \supset \,
\EL \hspace{0.02cm} \psiR \hspace{0.03cm} \etat ~,\quad
\psiRbar \hspace{0.02cm} \psiL ~,\quad
\Phi^\dag \eta \hspace{0.03cm} \chi^\ast / \Phi^\dag \eta \hspace{0.03cm} \chi^2 ~,\quad
\psiLcbar \psiL \chi ~,\quad
\chi^3 ~,~
+\,\text{h.c.} ~\,\,, \quad
\end{eqnarray}

\begin{table}[b]
\vspace{-0.2cm}
\begin{center}
\begin{tabular}
[c]{|c||c|c||c|c|c|c|c|}\hline
~$\vphantom{|_|^|}$~ & ~$E$~ &  ~$\Phi$~ &~$\psiL$~ & ~$\psiRc$~ &  ~$\eta$~ & ~$\chi$~ & ~$\sigma$~ 
\\\hline\hline
~SU(2)$_L\vphantom{|_|^|}$~   
& ~$\bm{2}$~ & ~$\bm{2}$~ & ~$\bm{1}$~ & $\bm{1}$ & ~$\bm{2}$~ & ~$\bm{1}$~& ~$\bm{1}$~ 
\\\hline
~U(1)$_{Y}\vphantom{|_|^|}$~ & ~$-$1/2~ & ~1/2~ & ~0~ & ~0~ & ~1/2~ & ~0~ & ~0~  
\\\hline
~U(1)$_{\text{B}-\text{L}}\vphantom{|_|^|}$~ 
& ~$-$1~ & ~0~ & ~$\QpsiL$~ & ~$\QpsiR$~ & ~$Q_\eta$~ & ~$Q_\chi$~ & ~$Q_\sigma$~
\\\hline
~Spin~$J\vphantom{|_|^|}$~ & ~1/2~ & ~0~ & ~1/2~ & ~1/2~ & ~0~& ~0~ & ~0~ 
\\\hline
\end{tabular}
\caption{
Charge assignments of the fermions and scalars in the two-loop $\UBL$ model II, 
where the unknown B$-$L charges can be fixed by using the same procedure.}
\label{tab:5}
\end{center}
\vspace{-0.8cm}
\end{table} 

\begin{figure}[t]
\begin{center}
\includegraphics[scale=0.25]{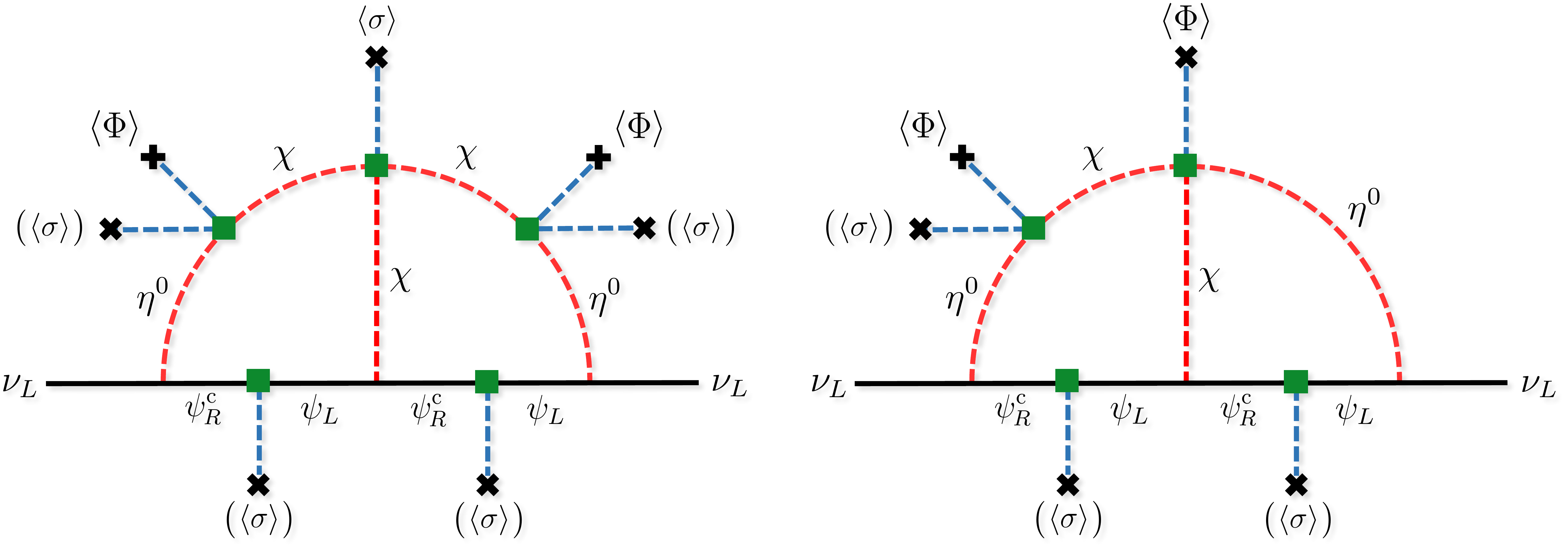}
\vspace{-0.2cm}
\caption{Feynman diagrams for the neutrino mass generation in the two-loop model II.}
\label{fig:two_loop_II}
\end{center}
\vspace{-0.3cm}
\end{figure}

Similarly to the first model, we introduce one singlet complex scalar $\sigma$ 
in order to break the $\UBL$ symmetry. 
The quantum numbers for new particles in the extended models are displayed in Table~\ref{tab:5}. 
The necessary interactions for the two-loop radiative seesaw mechanism are given by
\begin{eqnarray}
\mathcal{L} \,\, \supset
\begin{array}{l|l|l|ll|l}
& \psiRbar \hspace{0.03cm} \psiL 
& \Phi^\dag \eta \hspace{0.03cm} \chi^\ast 
& \psiLcbar \psiL \chi
& \psiRcbar \hspace{0.03cm} \psiR \hspace{0.03cm} \chi
& \chi^3 \sigma   \\
\EL \hspace{0.02cm} \psiR \hspace{0.03cm} \etat
& \psiRbar \hspace{0.03cm} \psiL \sigma
& \Phi^\dag \eta \hspace{0.03cm} \chi^\ast \sigma
& \psiLcbar \psiL \chi^\ast
& \psiRcbar \hspace{0.03cm} \psiR \hspace{0.03cm} \chi^\ast
& \chi^3 \sigma^\ast   \\
& \psiRbar \hspace{0.03cm} \psiL \sigma^\ast 
& \Phi^\dag \eta \hspace{0.03cm} \chi^\ast \sigma^\ast 
&&&
\end{array}
\hspace{-0.05cm}+\,\text{h.c.} ~.
\end{eqnarray}
At least one element in each column is needed for neutrino mass generation. There are two types of topological diagrams, as presented in Fig.~\ref{fig:two_loop_II}. 
By using the same prescription, we find nine possible charge assignments as shown in Table~\ref{tab:6}. 
The charge assignments C5, C6, and C9 reproduce the original $^{}\ZZZ$ model, 
while others forbid the Feynman diagram shown in the right panel of Fig.~\ref{fig:two_loop_II} 
since the term $\Phi^\dag\eta\chi^2$ is missing. 
%

\begin{table}[t]
\begin{center}
\begin{tabular}
[c]{|c||c|c|c|c|c||c|}\hline
$\vphantom{|_|^|}$ & ~$\QpsiL$~ & ~$\QpsiR$~ & ~$Q_\eta$~ & ~$Q_\chi$~ & ~$Q_\sigma$ & ~necessary interactions~ 
\\\hline\hline
~C1$\vphantom{|_|^|}$~ & ~$-$1/15~  & ~7/15~      & ~8/15~      & ~2/15~      & ~$-$2/5~  
&
~$
\EL \hspace{0.02cm} \psiR \hspace{0.03cm} \etat ~,~
\psiRbar \hspace{0.03cm} \psiL \sigma ~,~
\Phi^\dag \eta \hspace{0.03cm} \chi^\ast \sigma ~,~
\psiLcbar \psiL \chi ~,~
\chi^3 \sigma
$~
\\\hline
~C2$\vphantom{|_|^|}$~ & ~$-$1/9~    & ~1/9~        & ~8/9~        & ~2/9~        & ~$-$2/3~  
&
~$
\EL \hspace{0.02cm} \psiR \hspace{0.03cm} \etat ~,~
\psiRbar \hspace{0.03cm} \psiL ~,~
\Phi^\dag \eta \hspace{0.03cm} \chi^\ast \sigma ~,~
\psiLcbar \psiL \chi ~,~
\chi^3 \sigma
$~
\\\hline
~C3$\vphantom{|_|^|}$~ & ~$-$1/9~    & ~7/9~        &  ~2/9~       & ~2/9~        & ~$-$2/3~   
&
~$
\EL \hspace{0.02cm} \psiR \hspace{0.03cm} \etat ~,~
\psiRbar \hspace{0.03cm} \psiL \sigma ~,~
\Phi^\dag \eta \hspace{0.03cm} \chi^\ast ~,~
\psiLcbar \psiL \chi ~,~
\chi^3 \sigma
$~
\\\hline
~C4$\vphantom{|_|^|}$~ & ~$-$1/3~    & ~$-$5/3     & ~8/3~        & ~2/3~        & ~$-$2~      
&
~$
\EL \hspace{0.02cm} \psiR \hspace{0.03cm} \etat ~,~
\psiRbar \hspace{0.03cm} \psiL \sigma^\ast ~,~
\Phi^\dag \eta \hspace{0.03cm} \chi^\ast \sigma ~,~
\psiLcbar \psiL \chi ~,~
\chi^3 \sigma
$~
\\\hline
~C5$\vphantom{|_|^|}$~ & ~$-$1/3~    & ~7/3~        & ~$-$4/3~   & ~2/3~        & ~$-$2~      
&
~$
\EL \hspace{0.02cm} \psiR \hspace{0.03cm} \etat ~,~
\psiRbar \hspace{0.03cm} \psiL \sigma~,~
\Phi^\dag \eta \hspace{0.03cm} \chi^\ast \sigma^\ast ~,~
\Phi^\dag \eta \hspace{0.03cm} \chi^2 ~,~
\psiLcbar \psiL \chi ~,~
\chi^3 \sigma
$~
\\\hline
~C6$\vphantom{|_|^|}$~ & ~1/9~      & ~5/9~        & ~4/9~        & ~$-$2/9~   & ~$-$2/3~    
&
~$
\EL \hspace{0.02cm} \psiR \hspace{0.03cm} \etat ~,~
\psiRbar \hspace{0.03cm} \psiL \sigma ~,~
\Phi^\dag \eta \hspace{0.03cm} \chi^\ast \sigma ~,~
\Phi^\dag \eta \hspace{0.03cm} \chi^2 ~,~
\psiLcbar \psiL \chi ~,~
\chi^3 \sigma^\ast
$~
\\\hline
~C7$\vphantom{|_|^|}$~ & ~$-$1/3~    & ~1/3~        & ~2/3~        & ~2/3~         & ~$-$2~      
&
~$
\EL \hspace{0.02cm} \psiR \hspace{0.03cm} \etat ~,~
\psiRbar \hspace{0.03cm} \psiL~,~
\Phi^\dag \eta \hspace{0.03cm} \chi^\ast ~,~
\psiLcbar \psiL \chi ~,~
\chi^3 \sigma
$~
\\\hline
~C8$\vphantom{|_|^|}$~ & ~1/3~     & ~5/3~        & ~$-$2/3~   & ~$-$2/3~    & ~$-$2~      
&
~$
\EL \hspace{0.02cm} \psiR \hspace{0.03cm} \etat ~,~
\psiRbar \hspace{0.03cm} \psiL \sigma ~,~
\Phi^\dag \eta \hspace{0.03cm} \chi^\ast ~,~
\psiLcbar \psiL \chi ~,~
\chi^3 \sigma^\ast
$~
\\\hline
~C9$\vphantom{|_|^|}$~ & ~1/3~     & ~$-$1/3~   & ~4/3~        & ~$-$2/3~    & ~$-$2~      
&
~$
\EL \hspace{0.02cm} \psiR \hspace{0.03cm} \etat ~,~
\psiRbar \hspace{0.03cm} \psiL ~,~
\Phi^\dag \eta \hspace{0.03cm} \chi^\ast \sigma ~,~
\Phi^\dag \eta \hspace{0.03cm} \chi^2 ~,~
\psiLcbar \psiL \chi ~,~
\chi^3 \sigma^\ast
$~
\\\hline
\end{tabular}
\caption{
Possible charge assignments of the two-loop $\UBL\hspace{-0.05cm}$ model II. 
The last column contains the necessary interactions to generate neutrino masses at the two-loop level. 
}
\label{tab:6}
\end{center}
\vspace{-0.5cm}
\end{table}

\subsection{Three-loop $\mathbb{Z}_4$ Model}
\begin{figure}[b]
\begin{center}
\includegraphics[scale=0.32]{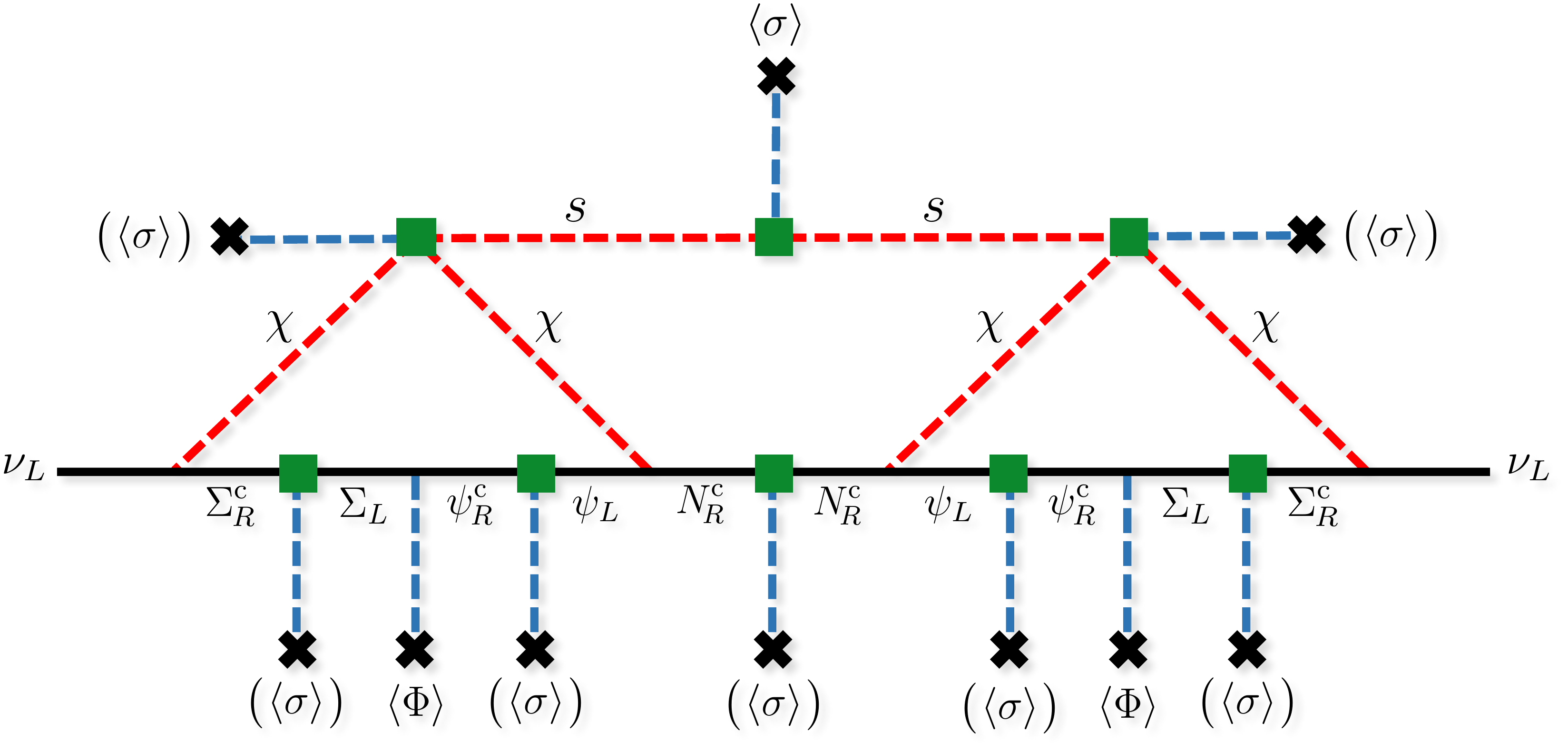}
\vspace{-0.2cm}
\caption{Feynman diagrams of the neutrino mass generation in the three-loop model.}
\label{fig:three_loop}
\end{center}
\vspace{-0.5cm}
\end{figure}

Based on the method we have developed, one can readily
build a radiative neutrino mass model with a remnant unbroken $\mathbb{Z}_{\text{N}}$ symmetry. 
The scalar interaction $\chi^{\text{N}}\sigma$ (or $\chi^{\text{N}}\sigma^\ast$) is needed 
to maintain the \,$\mathbb{Z}_{\text{N}}$ symmetry. 
%
As an example, we construct a three-loop radiative seesaw model with the $\UBL\hspace{-0.03cm}$ symmetry 
in which a dimensional-five operator $\chi^4\sigma$ (or $\chi^4\sigma^\ast$) is included in the diagram.
%
%
In order to make the model renormalizable, one more complex scalar field $s$ is introduced.
Through the trilinear scalar interactions $\chi^2s$ and $s^2\sigma$, 
the $\chi^4\sigma$ term can be generated when $s$ is integrated out. 
The particle contents and the charge assignments are defined in Table~\ref{tab:7}. 
The necessary interactions for generating neutrino masses at the three-loop level are 
\begin{eqnarray}
\mathcal{L}  \,\, \supset  
\begin{array}{c|l|l|l|l|ll|ll|l}
& \NRcbar \hspace{0.02cm} \NR
& \SigmaRbar \hspace{0.03cm} \SigmaL 
& \psiRbar \hspace{0.03cm} \psiL
& \SigmaLbar \psiR  \hspace{0.02cm} \Phit 
& \overline{\NR} \hspace{0.02cm} \psiL \chi
& \overline{N_{\hspace{-0.03cm}L}} \psiR \hspace{0.02cm} \chi
& \chi^2 s 
& \chi^2 s^\ast  
& s^2 \sigma \\
\EL \hspace{0.03cm} \SigmaR \hspace{0.03cm} \chi^\ast
&
\NRcbar \hspace{0.02cm} \NR \hspace{0.03cm} \sigma
& \SigmaRbar \hspace{0.03cm} \SigmaL \sigma
& \psiRbar \hspace{0.03cm} \psiL \sigma 
& \SigmaRbar \hspace{0.03cm}  \psiL \Phit
& \overline{\NR} \hspace{0.02cm} \psiL \chi^\ast
& \overline{N_{\hspace{-0.03cm}L}} \psiR \hspace{0.02cm} \chi^\ast
& \chi^2 s \hspace{0.02cm} \sigma
& \chi^2 s^\ast \sigma
& s^2 \sigma^\ast \\ 
&
\NRcbar \hspace{0.02cm} \NR \hspace{0.03cm} \sigma^\ast
& \SigmaRbar \hspace{0.03cm} \Sigma_L \sigma^\ast 
& \psiRbar \hspace{0.03cm} \psi_L \sigma^\ast
&&&
& \chi^2 s \hspace{0.02cm} \sigma^\ast
& \chi^2 s^\ast \hspace{0.02cm} \sigma^\ast
&\\
\end{array}
\nonumber\\[0.1cm]
+\,\,\text{h.c.} ~.
\end{eqnarray}
At least one element in each column is required to obtain the neutrino masses. One possible topological diagram for the neutrino mass generation is shown in Fig.~\ref{fig:three_loop}.\footnote{The other two possible topological diagrams can be obtained by different contractions of
$\hspace{0.02cm}\chi\hspace{0.02cm}$ in Fig.~\ref{fig:three_loop}. 
These diagrams must be added if one derives the complete neutrino mass formula for phenomenological discussions. Furthermore, depending on how the B$-$L charge is assigned, one may have accidental Yukawa interactions $\overline{N_{\hspace{-0.03cm}L}} \psiR \hspace{0.02cm} \chi^\ast$ (analogous to the assignments B5-B7) and/or a trilinear scalar coupling $\chi^2 s$  (similarly to the assignments C5, C6, and C9).
 } 
By integrating out the complex scalar field $s$, we arrive at the same topological diagram as the one in the KNT model~\cite{Krauss:2002px}. 
For a completeness, we give an example of the \hspace{0.01cm}B$-$L charge assignment in Table~\ref{tab:8}. 
All the possible assignments can be easily found by the same approach as mentioned above. After the $\UBL$ symmetry breaking, the remaining 
charges $\,(-1)^{\text{N}Q_{\text{B}-\text{L}}}$ under the discrete sym- metry ensure the stability of the  $\mathbb{Z}_4\hspace{-0.05cm}$ (non-trivially) charged particle in this model. The lightest $\mathbb{Z}_4\hspace{-0.05cm}$ charged particle is identified as a DM candidate, while $\NRc$ is the unique DM candidate in the KNT model. One interesting point is that this \,$\mathbb{Z}_4$ model may have an additional DM component if the decay of the heavier \,$\mathbb{Z}_4$ charged particle is kinematically forbidden. For example, in the case of the charge assignment in Table~\ref{tab:8}, the
particles $\NRc$ and $s$ have the \,$\mathbb{Z}_4$ charge $2$ while
$\Sigma$, $\psi$ and $\chi$ have different \,$\mathbb{Z}_4$ charges
than $2$. 
As a result, if the masses of the former particles are sufficiently light, their decay channels into the latter
particles are forbidden. 
Thus, in this case the \,$\mathbb{Z}_4$ model has two-component DM.

\begin{table}[t]
\begin{center}
\begin{tabular}
[c]{|c||c|c||c|c|c|c|c|c|c|c|}\hline
~$\vphantom{|_|^|}$~ & ~$E$~ &  ~$\Phi$~ & ~$\NRc$~ &~$\SigmaL$~ & ~$\SigmaRc$~ & ~$\psiL$~ & ~$\psiRc$~ & ~$\chi$~ & ~$s$~ & ~$\sigma$~ 
\\\hline\hline
~SU(2)$_L\vphantom{|_|^|}$~   
& ~$\bm{2}$~ & ~$\bm{2}$~ & ~$\bm{1}$~ & ~$\bm{2}$~ & $\bm{2}$ & ~$\bm{1}$~ & ~$\bm{1}$~& ~$\bm{1}$~ & ~$\bm{1}$~ & ~$\bm{1}$~ 
\\\hline
~U(1)$_{Y}\vphantom{|_|^|}$~ & ~$-$1/2~ & ~1/2~ & ~0~ & ~$-$1/2~ & ~1/2~ & ~0~ & ~0~ & ~0~ & ~0~ & ~0~ 
\\\hline
~U(1)$_{\text{B}-\text{L}}\vphantom{|_|^|}$~ 
& ~$-$1~ & ~0~ & ~$\QN$~ & ~$\QSigmaL$~ & ~$\QSigmaR$~ & ~$\QpsiL$~ & ~$\QpsiR$~ & ~$Q_\chi$~ & ~$Q_s$~ & ~$Q_\sigma$~
\\\hline
~Spin~$J\vphantom{|_|^|}$~ & ~1/2~ & ~0~ & ~1/2~ & ~1/2~ & ~1/2~ & ~1/2~& ~1/2~ & ~0~ & ~0~ & ~0~
\\\hline
\end{tabular}
\caption{
Charge assignments of the fermions and scalars in the three-loop $\UBL\hspace{-0.05cm}$ model.}
\label{tab:7}
\end{center}
\vspace{-0.2cm}
\end{table}

\begin{table}[t]
\begin{center}
\begin{tabular}
[c]{|c|c|c|c|c|c|c|c|}\hline
~~$\QN\vphantom{|_|^|}$~~ & ~$\QSigmaL$~ & ~$\QSigmaR$~ & ~$\QpsiL$~ & ~$\QpsiR$~ & ~$Q_\chi$~ & ~$Q_s$~ & ~$Q_\sigma$~           
\\\hline
~1/8~  & ~$-$9/16~ & ~13/16~ & ~$-$5/16~ & ~9/16~  & ~3/16~  & ~$-$1/8~ & ~$-$1/4~
\\\hline
\end{tabular}
\caption{
One possible B$-$L charge assignment of the new particles in the three-loop $\UBL\hspace{-0.05cm}$ model. 
The necessary interactions for generating the neutrino masses at three-loop level are 
$\EL \hspace{0.03cm} \SigmaR \hspace{0.03cm} \chi^\ast \,,\,
\NRcbar \NR \hspace{0.03cm} \sigma^\ast \,,\,
\SigmaRbar \hspace{0.03cm} \SigmaL \sigma \,,\,
\SigmaLbar \psiR  \hspace{0.02cm} \Phit  \,,\,
\overline{\NR} \hspace{0.02cm} \psiL \chi\,,\,
\chi^2 s  \hspace{0.02cm} \sigma$ and $s^2 \sigma^\ast.$}
\label{tab:8}
\end{center}
\vspace{-0.5cm}
\end{table}

\section{Discussion : Anomaly Cancellation}
\label{sec:3}
All of the models we have discussed so far give rise to gauge anomalies which should not appear in a consistent gauge theory at the quantum level. However, these anomalies can be canceled by introducing new exotic fermions~\cite{Appelquist:2002mw, Batra:2005rh, Kanemura:2011mw}. 
For most of the models described above (strictly speaking for the one-loop $\ZZ$ model, 
and the B2, B7, B9 assignments of the two-loop $\ZZZ$ model I, and all the assignments of the two-loop $\ZZZ$ model II,  
it is sufficient to impose only two kinds of anomaly cancellation conditions for [gravity]$^2\hspace{-0.03cm}\otimes\hspace{-0.03cm}\UBL\hspace{-0.03cm}$ and $[\UBL]^3$ 
since all the new fermions in these models are the vector-like pairs and/or the charge neutral under the SM gauge group. However, for the other cases, since some pairs of the SU(2)$_L\hspace{-0.05cm}$ doublet fermions are included in the particle content, the anomaly cancellation conditions become complicated. Three more conditions for [SU(2)$_L]^2\hspace{-0.05cm}\otimes\hspace{-0.05cm}\UBL^{}$,
$[\UBL]^2\hspace{-0.05cm}\otimes\hspace{-0.05cm}\text{U(1)}_Y$ and 
$\UBL\hspace{-0.05cm}\otimes\hspace{-0.05cm}[\text{U(1)}_Y]^2$ 
are required. Nevertheless, all the anomalies can basically be canceled by adding new exotic fermions. We give here an example of the anomaly cancellations for the one-loop $\ZZ$ model and the two-loop $\ZZZ$ model II. 
\begin{table}[b]
\begin{center}
\begin{tabular}{|c||c|c|c|c|c|}\hline
$\vphantom{|_|^|}$ & $\xi_L$ & $\xi_R^\text{c}$ & $\zeta_L$ & $\zeta_R^\text{c}$ & $\lambda_L$\\\hline\hline
$\UBL\vphantom{|_|^|}$ 
& ~\,$Q_{\xi_L}$~ & ~\,$Q_{\xi_R}$~ & ~\,$Q_{\zeta_L}$~ 
& ~\,$Q_{\zeta_R}$~ & ~$Q_{\lambda_L}$~ \\\hline
~Generations$\vphantom{|_|^|}$~ 
& \multicolumn{2}{|c|}{$n_\xi$} & \multicolumn{2}{|c|}{$n_\zeta$} & $n_\lambda$\\\hline
\end{tabular}
\caption{A list of exotic fermions added for anomaly cancellations. 
Their B$-$L charges and the number of generations are defined.}
\label{tab:fermion_anomaly}
\end{center}
\vspace{-0.5cm}
\end{table}

For the anomaly cancellations, we introduce $n_\xi$ pairs of Dirac fermions $(\xi_L,\,\xi_R^\text{c})$\,, 
$n_\zeta$ pairs Dirac fermions of $(\zeta_L,\,\zeta_R^\text{c})$, and $^{}n_\lambda$ generations of Majorana fermions $^{}\lambda_L$ whose B$-$L charges are defined in Table~\ref{tab:fermion_anomaly}, 
where all the new fermions transform as singlets under the SM gauge group. In order to assign nonzero $\UBL\hspace{-0.04cm}$ charges for exotic fermions while keeping the non-vector-like nature under the $\UBL\hspace{-0.03cm}$ symmetry, the masses of these fermions are induced by the VEV $\langle\sigma\rangle$.
To be specific, $\UBL\hspace{-0.03cm}$ charges of the exotic fermions are fixed by the terms
$\sigma^\ast\overline{\xi_R}\hspace{0.03cm}\xi_L$\,,\,$\sigma\hspace{0.03cm}\overline{\zeta_R}\hspace{0.03cm}\zeta_L$ 
and $\sigma^\ast\overline{\lambda_L^\text{c}}\lambda_L$. 
Theseareencodedtothe following relations : $\hspace{-0.1cm}Q_{\xi_L}+\,Q_{\xi_R} = Q_\sigma$\,,\,
$Q_{\zeta_L}+\,Q_{\zeta_R} = -\, Q_\sigma$ and $2^{}Q_{\lambda_L}\hspace{-0.07cm}=Q_\sigma$. 
We would like to make one more point, namely that the charges of the exotic fermions have to be chosen so as not to affect the aforementioned models. For instance, we impose $Q_{\text{B}-\text{L}}\neq1$ for the
exotic fermions $\xi$, $\zeta$ and $\lambda$; 
otherwise, the symmetry allows the tree-level neutrino Yukawa coupling, which generates the neutrino masses by the canonical seesaw mechanism.

The conditions of anomaly cancellations for [gravity]$^2\hspace{-0.05cm}\otimes\hspace{-0.05cm}\UBL\hspace{-0.03cm}$ 
and $[\UBL]^3$ are given by
\begin{eqnarray}
\sum \hspace{-0.05cm} Q_{\text{B}-\text{L}} \,=\,
\sum \hspace{-0.05cm} Q_{\text{B}-\text{L}}^3 \,=\, 0 ~,
\end{eqnarray}
where the summation is taken over all the fermions included in a model. For the one-loop $\ZZ$ model and two-loop $\ZZZ$ model II, these conditions are explicitly given by
\begin{eqnarray}
-^{}3^{}+\sum \hspace{-0.05cm} Q_\text{model} {}^{} + 
\scalebox{1.2}{\big(}n_\xi-n_\zeta +
\tfrac{1}{2}^{}n_\lambda
\scalebox{1.2}{\big)}^{} Q_\sigma \,=\, 0 ~,
\end{eqnarray}
\vspace{-0.7cm}
\begin{eqnarray}
-^{}3^{}+\sum  \hspace{-0.05cm} Q_\text{model}^3 {}^{} +
\scalebox{1.2}{\big[}\big(n_\xi-n_\zeta\big)+\tfrac{1}{8}n_\lambda\scalebox{1.2}{\big]}Q_\sigma^3
-3^{} \scalebox{1.2}{\big(}
n_\xi{}^{}Q_{\xi_L}\hspace{-0.05cm} Q_{\xi_R} - n_{\zeta}{}^{}Q_{\zeta_L} \hspace{-0.05cm} Q_{\zeta_R}
\scalebox{1.2}{\big)} Q_{\sigma} \,=\, 0~ ,
\end{eqnarray}
where $\sum \hspace{-0.05cm} Q_\text{model}$ and $\sum \hspace{-0.05cm} Q_\text{model}^3$ 
represent contributions from the new fermions in each radiative seesaw model. Those from the exotic fermions for the anomaly cancellation are separately taken into account. For each model class, we have
\begin{eqnarray}
\sum \hspace{-0.05cm} Q_\text{model} \,=\, \left\{
\begin{array}{cl}
n_{\hspace{-0.02cm}N} \hspace{0.02cm} \QN  &\quad \text{for one-loop $\ZZ$ model} \\
n_\psi\big(\QpsiL+\QpsiR\big) &\quad \text{for two-loop $\ZZZ$ model II}
\end{array}
\right.~,
\end{eqnarray}
and
\begin{eqnarray}
\sum \hspace{-0.05cm} Q_\text{model}^3 \,=\, \left\{
\begin{array}{cl}
n_{\hspace{-0.02cm}N} \hspace{0.02cm} \QN^3 & \quad \text{for one-loop $\ZZ$ model}\\
n_\psi\big(\QpsiL^3+\QpsiR^3\big) & \quad \text{for two-loop $\ZZZ$ model II}
\end{array}
\right.~,
\end{eqnarray}
where the number of generations of $N$ and $\psi$ represented 
by $^{}n_{\hspace{-0.02cm}N}$ and $^{}n_\psi\,$, respectively. 
The $\UBL$ charges $\QN$, $\QpsiL\hspace{-0.05cm}$ and $^{}\QpsiR$ are listed 
in Table~\ref{tab:2} and \ref{tab:6} for each model. 
By solving the simultaneous equations, one can find a set of solutions satisfying these conditions. 
Example sets of \,B$-$L charge assignments and numbers of generations being consistent with 
the gauge anomalies in each model are given in Table~\ref{tab:anomaly1} and \ref{tab:anomaly3} in Appendix. 

Even if we add these new exotic fermions for anomaly cancellations, the discussion of the remnant discrete symmetry from the gauge symmetry is kept unchanged because the exotic fermions are introduced in a sector that is completely separate from these models. We should like to make one more comment regarding the new exotic fermions. After the spontaneous breaking of the $\UBL$ symmetry, different discrete symmetries can accidentally appear in this new sector at the renormalizable level. This means that new DM candidates emerge and thus the models have multi-component DM, which can interact with each other through the $Z^\prime\hspace{-0.06cm}$ and Higgs bosons. The appearance of additional DM candidates is a rather common feature of the extended models. One might think that some specific choice of B$-$L charges for exotic fermions can allow the Yukawa interactions with the first DM sector. However, in all the models classified in the one-loop $\ZZ$ model and in the two-loop $\ZZZ$ model II, 
we confirm that any choice of B$-$L charges for $\xi^{}, \zeta$ and $\lambda$ leads the second DM candidate. 
It is impossible to allow Yukawa interactions for all exotic fermions even through the mass mixing among them. A possible way to avoid a multi-component DM scenario is to introduce the new scalar fields, which connect the new exotic fermions to the other fermions in the model. In such a case, the VEVs of new scalar fields can break unwanted accidental discrete symmetries.\footnote{We have to choose the charges of new exotic fermions that do not disturb the stability of the true DM particle.}

As for the multi-component DM scenario, since the masses of the exotic fermions are generated by the VEV $\langle\sigma\rangle$, 
the mass scale of most of the new particles is roughly expected to be of the same order.\footnote{In some of the models, the masses of fermions are provided by explicit mass terms without the VEV $\langle\sigma\rangle$.}
When the relic density of DM is calculated, the number density of multi-component DM can be changed by conversion processes. Therefore, one has to solve the coupled Boltzmann equations for multi-component DM in order to properly compute the DM relic abundance
~\cite{Belanger:2012vp, Aoki:2012ub, Belanger:2014vza}. 
Since the exotic fermions interact only with the $Z^\prime$ gauge boson and the Higgs bosons via the mixing with $\sigma$ at the tree level, the main annihilation mode of additional DM components would be described by these interactions. The other DM component originated by the remnant symmetry of the $\UBL\hspace{-0.02cm}$ symmetry has the Yukawa and scalar interactions, which are relevant to the neutrino mass generation sector. Thus, the fraction of the relic abundance for each DM component would be determined by the relative strengths of these couplings.

Moreover, the new exotic fermions may associate with other aspects of DM phenomenology such as detection properties. For example, for direct detection of DM, the elastic scattering with nuclei can be induced by the $t$-channel process mediated by $Z^\prime$ gauge boson and the Higgs bosons for all of the DM components. The detection rate for each multi-component DM would be similar unless hierarchical coupling constants are considered. The recoil-energy spectrum for elastic scattering with nuclei can be a discriminant if the masses of the multiple DM compo- nents are nondegenerate~\cite{Profumo:2009tb, Dienes:2012cf}. 
Furthermore, in models with multi-component DM, the multiple monochromatic gamma-ray or the multiple neutrino lines at distinct energies can generally be predicted as indirect detection signals due to the mass splitting among multiple DM components.\footnote{
Such a multiple peak may not be a clear signal of multi-component DM since it is possible to generate similar gamma-rays or neutrino lines even in the single-component DM case through different annihilation channels~\cite{Bertone:2009cb}. }
In the above models, all of the DM components can basically generate the monochromatic gamma-ray spectrum due to the loop-induced two-body annihilation channel into $\gamma\gamma$ through the interactions with the electromagnetically charged particles in the SM and the new sector. More specifically, since the first DM component has the Yukawa interactions (which are required to generate the neutrino masses) in addition to the U(1)$_{\text{B}-\text{L}}$ gauge interaction, a stronger monochro- matic gamma-ray signal would be induced. The signal strength would be much higher than that of the other DM components made of the lightest exotic fermion for anomaly cancellations. A detailed exploration of the phenomenology of multi-component DM is beyond the scope of this paper, and it will be discussed elsewhere.

\section{Summary and Conclusions}
\label{sec:5}
We have presented a prescription for classifying the gauged $\UBL\hspace{-0.05cm}$ models for the radiative neutrino mass
B$-$L
generation and the DM stability. In this class of models, the
tiny neutrino masses are naturally explained by the loop suppression of the radiative seesaw mechanism, while the DM stability is automatically maintained by the residual symmetry of the spontaneous $\UBL$ symmetry breaking. These models are systematically classified by the identifications (the insertion of the VEV) of the B$-$L breaking vertices in the prototype models for the loop-induced neutrino masses with a discrete symmetry. We found five independent models for the one-loop $\ZZ$ model, and nine independent models for each two-loop \,$\ZZZ$\, model. This procedure is easily extended to the models based on higher loop diagrams and \,$\ZN$\, symmetry. These minimal models generally contain gauge anomalies, which can be easily canceled by introducing the exotic fermions. Since additional discrete symmetries appear in the exotic fermion sector, these models tend to have multiple DM candidates. 


\section*{Acknowledgments}
T.T. acknowledges support from P2IO Excellence Laboratory (LABEX). K.T.'s work is supported in part by the MEXT Grant-in-Aid for Scientific Research on Innovative Areas No. 26104704, and No. 16H00868, the JSPS Grant-in-Aid for Young Scientists (B) No. 16K17697, and the Supporting Program for Interaction-based Initiative Team Studies (Kyoto University).


\section*{Appendix: Fermion Contents and Charges for Anomaly Cancellations}
An example of B$-$L charge assignments for anomaly cancellations is given 
in Table~\ref{tab:anomaly1} for the one-loop $\ZZ$ model and 
in Table~\ref{tab:anomaly3} for the two-loop $\ZZZ$ model II. 

\begin{table}[h]
\vspace{0.5cm}
\begin{center}
\begin{tabular}{|c||c||c|c|c|c|c||c|c|c|c|}\hline
$\vphantom{|_|^|}$ & $\QN$ & $Q_{\xi_L}$ & $Q_{\xi_R}$ & $Q_{\zeta_L}$ & $Q_{\zeta_R}$ & $Q_{\lambda_L}$ &
~$n_{\hspace{-0.02cm}N}$~ & ~$n_\xi$~ & ~$n_\zeta$~ & ~$n_\lambda$~  \\\hline\hline
~A1$\vphantom{|_|^|}$~ & ~1/2~ & ~1/10~  & ~$-$11/10~ & ~6/5~ & ~$-$1/5~
		     & ~$-\hspace{-0.2cm}-$~ & ~4~ & ~2~ & ~3~ & ~$-\hspace{-0.2cm}-$~ \\\hline
~A2$\vphantom{|_|^|}$~ & ~1/4~ & ~$-$1/8~ & ~$-$3/8~  & ~15/16~ & ~$-$7/16~ & ~$-\hspace{-0.2cm}-$~ & ~6~ & ~1~ & ~4~ & ~$-\hspace{-0.2cm}-$~ \\\hline
~A3$\vphantom{|_|^|}$~ & ~$-$1/2~ & ~1/6~ & ~5/6~ & ~$-$3/4~ & ~$-$1/4~ & ~$-\hspace{-0.2cm}-$~ & ~4~ & ~9~ & ~4~ & ~$-\hspace{-0.2cm}-$~ \\\hline
~A4$\vphantom{|_|^|}$~ & ~0~ & ~7/3~ & ~$-$13/3~ & ~8/3~ & ~$-$2/3~ & ~$-$1~ & ~3~ & ~1~ & ~4~ & ~3~ \\\hline
~A5$\vphantom{|_|^|}$~ & ~0~ & ~$-$1/9~ & ~$-$5/9~ & ~$-$1/9~ & ~7/9~ & ~$-$1/3~ & ~3~ & ~1~ & ~7~ & ~3~ \\\hline
\end{tabular}
\caption{Charge assignments and the number of generations required for the anomaly cancellation in the one-loop $\ZZ$ model.}
\label{tab:anomaly1}
\end{center}
\end{table}

\begin{table}[h]
\begin{center}
\begin{tabular}{|c||c|c||c|c|c|c|c||c|c|c|c|}\hline
$\vphantom{|_|^|}$ & $\QpsiL$ & $\QpsiR$ & $Q_{\xi_L}$ & $Q_{\xi_R}$ & $Q_{\zeta_L}$ & $Q_{\zeta_R}$ & $Q_{\lambda_L}$ & ~$n_\psi$~ & ~$n_\xi$~ & ~$n_\zeta$~ & ~$n_\lambda$~ \\\hline\hline
~C1$\vphantom{|_|^|}$~ & 
~$-$1/15~ & ~7/15~ & ~0~ & ~$-$2/5~ & ~11/15~ & ~$-$1/3~ & ~$-$1/5~ & ~4~ & ~4~ & ~8~ & ~1~ \\\hline
~C2$\vphantom{|_|^|}$~ &
~$-$1/9~ & ~1/9~ & ~$-$1/6~ & ~$-$1/2~ & ~3/4~ & ~$-$1/12~ & ~$-$1/3~ & ~3~ & ~2~ & ~8~ & ~3~ \\\hline
~C3$\vphantom{|_|^|}$~ & 
~$-$1/9~ & ~7/9~ & ~2/3~ & ~$-$4/3~ & ~11/9~ & ~$-$5/9~ & ~$-$1/3~ & ~3~ & ~4~ & ~6~ & ~1~ \\\hline
~C4$\vphantom{|_|^|}$~ & 
~$-$1/3~ & ~$-$5/3~ & ~1/6~ & ~$-$13/6~ & ~7/4~ & ~1/4~ & ~$-$1~ & ~4~ & ~2~ & ~8~ & ~1~ \\\hline
~C5$\vphantom{|_|^|}$~ & 
~$-$1/3~ & ~7/3~ & ~4/3~ & ~$-$10/3~ & ~4/3~ & ~2/3~ & ~$-$1~ & ~3~ & ~1~ & ~1~ & ~3~ \\\hline
~C6$\vphantom{|_|^|}$~ & 
~1/9~ & ~5/9~ & ~11/9~ & ~$-$17/9~ & ~5/3~ & ~$-$1~ & ~$-$1/3~ & ~4~ & ~1~ & ~2~ & ~1~ \\\hline
~C7$\vphantom{|_|^|}$~ & 
~$-$1/3~ & ~1/3~ & ~11/3~ & ~$-$17/3~ & ~10/3~ & ~$-$4/3~ & ~$-$1~ & ~3~ & ~1~ & ~4~ & ~3~ \\\hline
~C8$\vphantom{|_|^|}$~ & 
~1/3~ & ~5/3~ & ~7/3~ & ~$-$13/3~ & ~5~ & ~$-$3~ & ~$-$1~ & ~3~ & ~3~ & ~2~ & ~1~ \\\hline
~C9$\vphantom{|_|^|}$~ & 
~1/3~ & ~$-$1/3~ & ~5~ & ~$-$7~ & ~4~ & ~$-$2~ & ~$-$1~ & ~3~ & ~1~ & ~4~ & ~3~ \\\hline
\end{tabular}
\caption{Charge assignments and the number of generations required for the anomaly cancellation in the two-loop $\mathbb{Z}_3$ model II.}
\label{tab:anomaly3}
\end{center}
\end{table}


\end{document}